\documentclass[nofootinbib,prd,twocolumn,showpacs,showkeys,preprintnumbers]{revtex4-1}
\usepackage{hyperref,amssymb,amsmath,mathrsfs,bm,graphicx}
\begin{document}

\title {Earliest stages of the non--equilibrium in axially symmetric, self-gravitating,  dissipative fluids}
\author{L. Herrera}
\email{lherrera@usal.es}
\affiliation{Escuela de F\'\i sica, Facultad de Ciencias, Universidad Central de Venezuela, Caracas, Venezuela and Instituto Universitario de F\'isica
Fundamental y Matem\'aticas, Universidad de Salamanca, Salamanca, Spain}
\author{A. Di Prisco}
\email{alicia.diprisco@ciens.ucv.ve}
\affiliation{Escuela de F\'\i sica, Facultad de Ciencias, Universidad Central de Venezuela, Caracas, Venezuela}
\author{J. Ospino}
\email{j.ospino@usal.es}
\affiliation{Departamento de Matem\'atica Aplicada and Instituto Universitario de F\'isica
Fundamental y Matem\'aticas, Universidad de Salamanca, Salamanca, Spain}
\author{J.Carot }
\email{jcarot@uib.cat}
\affiliation{Departament de  F\'{\i}sica, Universitat Illes Balears, E-07122 Palma de Mallorca, Spain}
\begin{abstract}
We report   a study on  axially and reflection symmetric dissipative fluids,
just after its departure from hydrostatic and thermal  equilibrium, at  the smallest  time 
scale at which the first signs of dynamic evolution appear. Such  a time scale   is smaller than the thermal relaxation time, the thermal adjustment time and the hydrostatic time. It is obtained 
that the onset of non--equilibrium will critically depend on a  single function directly related to the time derivative of the vorticity.  Among all fluid variables (at the time scale under consideration),  only the  tetrad component  of  the anisotropic tensor  in the subspace orthogonal to the four--velocity and the Killing vector of axial symmetry, shows signs of dynamic evolution. Also, the first step toward a dissipative regime begins with a non--vanishing time derivative of the heat flux component along the meridional direction. The magnetic part of the Weyl tensor vanishes (not so its time derivative), indicating that the emission of gravitational radiation will occur at later times. Finally, the decreasing of the effective inertial mass density, associated to thermal effects, is clearly illustrated.
\end{abstract}
\maketitle

\section{Introduction}
Many issues related with the structure of self--gravitating fluids may be addressed within  the static regime.
In this   case, the spacetime admits a timelike, hypersurface orthogonal, Killing vector. Thus, a coordinate system can always be chosen, such that all metric and physical variables are independent on the time like coordinate. The static case, for axially and reflection symmetric spacetimes, was studied in \cite{static}. In such a case the fluid is in equilibrium, implying that the hydrostatic equilibrium equations (Eqs.(21,22) in \cite{static}) are satisfied.

If, instead,   the system evolves with time,  we have to consider  the full dynamic case where the system is  out of equilibrium (thermal and dynamic), the general formalism to analyze this situation, for axially  and reflection symmetric spacetimes was developed in \cite{1} using a framework based on the $1+3$ formalism \cite{21cil, n1, 22cil, nin}.

However, some  part of the life of stars (at any stage of evolution), may be 
described on the basis of the quasi-static approximation (slowly 
evolving regime). This is so, because many relevant processes in star 
interiors take place on time scales that are usually, much larger 
than the hydrostatic time scale \cite{astr1},\cite{astr2}. In this case, the system is assumed to evolve, although  slowly enough, so   that  the hydrostatic equilibrium equations (Eqs.(21,22) in \cite{static}) are assumed to be satisfied, all along the evolution.

This regime   has been recently described in detail, within the context of the $1+3$ formalism \cite{qsa}.
\noindent

Nevertheless, during their evolution, self-gravitating objects may pass 
through phases of intense dynamical activity for which the quasi-static 
approximation is clearly not reliable (e.g., the quick collapse phase 
preceding neutron star formation). 

It is worth mentioning that both regimes (``quick'' and ``quasi--static''), may be present, at different phases of the collapse of massive stars. Indeed, after the core bounce, leading to a supernova, the hydrostatic equilibrium is reached within few milliseconds, while the subsequent, Kelvin--Helmholtz  phase, lasts for about 20 seconds, during which the system is in the quasi--static regime, thereby satisfying the hydrostatic equilibrium equations \cite{burr}. We recall, that the hydrostatic time for a neutron star is of the order of $10^{-3}$ seconds, while the order of magnitude of the relaxation time for neutron star matter range from $10^{-3}$ to $10^{-1}$ seconds.

All these phases of star evolution 
(``slow'' and ``quick'') are generally accompanied by intense dissipative 
processes, usually described in the diffusion approximation. 

This 
assumption, in its turn, is justified by the fact that frequently, 
the mean free path of particles responsible for the propagation of 
energy in stellar interiors is very small as compared with the typical 
length of the star.

\noindent
Here we shall  focus on the ``quick'' phase, with the inclusion of all the dissipative processes.

\noindent
However, instead of following the  evolution of the system for a long time 
after its departure from equilibrium, we 
shall analyze its behaviour immediately after such departure.

In this work  ``immediately'' means at the smallest  time scale, at which we can observe the first signs of dynamical evolution.  Such a time scale is assumed to be  smaller than the thermal relaxation time, the hydrostatic time, and the thermal adjustment time.

\noindent
Doing so we shall be able to extract important conclusions about  the very early stages of non--equilibrium, avoiding the introduction of numerical procedures  
which might lead to model dependent conclusions.

\noindent
The price to pay for such a simplification, is that we shall  describe only  the very early stages of the evolution. The reward is that we shall be able to answer to the following questions:
\begin{enumerate}
\item what are the first signs of non--equilibrium?
\item  what physical variables do exhibit such signs?
\item what does control the onset of the dynamic regime, from an equilibrium initial configuration?
\end{enumerate}

Our approach may be summarized as follows: We observe a system which is initially static, and leaves the equilibrium for unknown causes which are not relevant for the discussion. At this moment we put the clock to work, and watch the system until the first signs of non--equilibrium appear. At this very moment, we stop the clock. It is during this time scale that we describe the behaviour of the system
\noindent

As we shall see, a specific function related  with the time derivative of the vorticity vector, appears as the fundamental variable, controlling the departure from equilibrium and the ensuing evolution. By  analogy (in its physical meaning) with the Bondi's news function \cite{7}, we shall refer to this quantity as the fluid news function.

From the analysis of the transport equation we shall see that the time derivative of one of the heat flux components (``radial'')  vanishes at the time scale under consideration, whereas the time derivative of the other (``meridional'') component, is controlled by the fluid news function. 

Also we shall see that, at the time scale under consideration,  the only fluid variable which exhibits deviation from the equilibrium is the tetrad component of the anisotropic tensor  in the subspace spanned by the two space--like vectors orthogonal to the four--velocity and the Killing vector of axial symmetry.

At this same time scale, the magnetic part of the Weyl tensor vanishes, implying that no emission of gravitational radiation is produced at this stage of evolution.  However, the time derivative of the magnetic part of the Weyl tensor does not vanish and depends upon the fluid news function, in such a way, that the vanishing of the latter imply the vanishing of the former. In other words the emission of gravitational process occurs at a time scale larger than the one considered here, and is tightly related to the fluid news function.

Finally, by using the transport equations together with the ``conservation'' laws, we put in evidence the decreasing of the effective inertial mass density, associated with thermal effects. 

In this work we shall heavily rely on the formalism developed in \cite{1}, thus in order to avoid the rewriting of some of the equations we shall frequently refer to \cite{1}, however  we warn  the reader of some important changes in the notation.
\section{Basic definitions and notation}
In this section we shall deploy all the variables required for our study, some details of the calculations are given  in \cite{1}, and therefore we shall omit them here.
\subsection{The metric, the source, and the kinematical variables}

We shall consider,  axially (and reflection) symmetric sources. For such a system the  line element may be written in ``Weyl spherical coordinates'' as:

\begin{equation}
ds^2=-A^2 dt^2 + B^2 \left(dr^2
+r^2d\theta^2\right)+C^2d\phi^2+2Gd\theta dt, \label{1b}
\end{equation}
where $A, B, C, G$ are positive functions of $t$, $r$ and $\theta$. We number the coordinates $x^0=t, x^1=r, x^2= \theta, x^3=\phi$.

We shall assume that  our source is filled with an anisotropic and dissipative fluid. We are concerned with either bounded or unbounded configurations. In the former case we should further assume that the fluid is bounded by a timelike surface $S$, and junction (Darmois) conditions should be imposed there.

The energy momentum tensor may be written in the ``canonical'' form, as 
\begin{equation}
{T}_{\alpha\beta}= (\mu+P) V_\alpha V_\beta+P g _{\alpha \beta} +\Pi_{\alpha \beta}+q_\alpha V_\beta+q_\beta V_\alpha.
\label{6bis}
\end{equation}

The above is the canonical, algebraic decomposition of a second order symmetric tensor with respect to unit timelike vector, which has the standard physical meaning when $T_{\alpha \beta}$ is the energy-momentum tensor describing some energy distribution, and $V^\mu$ the four-velocity assigned by certain observer.

With the above definitions it is clear that $\mu$ is the energy
density (the eigenvalue of $T_{\alpha\beta}$ for eigenvector $V^\alpha$), $q_\alpha$ is the  heat flux, whereas  $P$ is the isotropic pressure, and $\Pi_{\alpha \beta}$ is the anisotropic tensor. We emphasize that we are considering an Eckart frame  where fluid elements are at rest.

Since we choose the fluid to be comoving in our coordinates, then
\begin{equation}
V^\alpha =\left(\frac{1}{A},0,0,0\right); \quad  V_\alpha=\left(-A,0,\frac{G}{A},0\right).
\label{m1}
\end{equation}

We shall next define a canonical  orthonormal tetrad (say  $e^{(a)}_\alpha$), by adding to the four--velocity vector $e^{(0)}_\alpha=V_\alpha$, three spacelike unitary vectors (these correspond to the vectors $\bold K, \bold L, \bold S$ in \cite{1})

\begin{equation}
e^{(1)}_\alpha=(0,B,0,0); \quad  e^{(2)}_\alpha=\left(0,0,\frac{\sqrt{A^2B^2r^2+G^2}}{A},0\right),
\label{7}
\end{equation}

\begin{equation}
 e^{(3)}_\alpha(0,0,0,C),
\label{3nb}
\end{equation}
with $a=0,\,1,\,2,\,3$ (latin indices labeling different vectors of the tetrad)

The  dual vector tetrad $e_{(a)}^\alpha$  is easily computed from the condition 
\begin{equation}
 \eta_{(a)(b)}= g_{\alpha\beta} e_{(a)}^\alpha e_{(b)}^\beta, \qquad e^\alpha_{(a)}e_\alpha^{(b)}=\delta^{(b)}_{(a)},
\end{equation}
where $\eta_{(a)(b)}$ denotes the Minkowski metric.

In the above, the tetrad vector $e_{(3)}^\alpha=(1/C)\delta^\alpha_\phi$ is parallel to
the only admitted Killing vector (it is the unit tangent to the orbits of the
group of 1--dimensional rotations that defines axial symmetry). The other two
basis vectors $e_{(1)}^\alpha,\,e_{(2)}^\alpha$ define the two {\it unique}
directions that are orthogonal to the 4--velocity and to the Killing vector.

For the energy density and the isotropic pressure, we have
\begin{equation}
\mu=T_{\alpha \beta}e^\alpha_{(0)}e^\beta_{(0)},\qquad P=\frac{1}{3}h^{\alpha \beta}T_{\alpha \beta},
\label{eisp}
\end{equation}
where
\begin{equation}
h^\alpha_{\beta}=\delta ^\alpha_{\beta}+V^\alpha V_{\beta},
\label{vel5}
\end{equation}
whereas the anisotropic tensor  may be  expressed through three scalar functions defined as (see \cite{1}, but notice the change of notation):

\begin{eqnarray}
 \Pi _{(2)(1)}=e^\alpha_{(2)}e^\beta_{(1)} T_{\alpha \beta} 
, \quad , \label{7P}
\end{eqnarray}

\begin{equation}
\Pi_{(1)(1)}=\frac{1}{3}\left(2e^{\alpha}_{(1)} e^{\beta}_{(1)} -e^{\alpha}_{(2)} e^{\beta}_{(2)}-e^{\alpha}_{(3)} e^{\beta}_{(3)}\right) T_{\alpha \beta},
\label{2n}
\end{equation}
\begin{equation}
\Pi_{(2)(2)}=\frac{1}{3}\left(2e^{\alpha}_{(2)} e^{\beta}_{(2)} -e^{\alpha}_{(3)} e^{\beta}_{(3)}-e^{\alpha}_{(1)} e^{\beta}_{(1)}\right) T_{\alpha \beta}.
\label{2nbis}
\end{equation}

This specific choice of  these scalars is justified by the fact, that  the relevant equations used  to carry out this study,  become more compact and easier to handle, when expressed in terms of  them.

Finally, we may write the heat flux vector in terms of  the two tetrad components $q_{(1)}$ and $q_{(2)}$:
\begin{equation}
q_\mu=q_{(1)}e_{\mu}^{(1)}+q_{(2)}e_{\mu}^{(2)}
\label{qn1}
\end{equation}
or, in coordinate components (see \cite{1})
\begin{equation}
q^\mu=\left(\frac{q_{(2)} G}{A \sqrt{A^2B^2r^2+G^2}},  \frac{q_{(1)}}{B}, \frac{Aq_{(2)}}{\sqrt{A^2B^2r^2+G^2}}, 0\right)
,\label{q}
\end{equation}
\begin{equation}
 q_\mu=\left(0, B q_{(1)}, \frac{\sqrt{A^2B^2r^2+G^2}q_{(2)}}{A}, 0\right).
\label{qn}
\end{equation}

Of course, all the above quantities depend,  in general, on $t, r, \theta$.

The expressions for the kinematical variables are  (see  \cite{1}).

For the four acceleration we have
\begin{equation}
a_\alpha=V^\beta V_{\alpha;\beta}=a_{(1)}e_{\mu}^{(1)}+a_{(2)}e_{\mu}^{(2)},
\label{a1n}
\end{equation}
with
\begin{equation}
a_{(1)}= \frac {A^\prime }{AB };\quad a_{(2)}=\frac{A}{\sqrt{A^2B^2r^2+G^2}}\left[\frac {A_{,\theta}}{A}+\frac {G}{A^2}\left(\frac{\dot G}{G}-\frac{\dot A}{A}\right)\right],
\label{acc}
\end{equation}
 where the dot  and the prime denote derivatives with respect to $t$ and $r$ respectively. 

For the expansion scalar
\begin{eqnarray}
\Theta&=&V^\alpha_{;\alpha}=\frac{1}{A}\left(\frac{2 \dot B}{B}+\frac{\dot C}{C}\right) \nonumber\\
&+&\frac{G^2}{A\left(A^2 B^2 r^2 + G^2\right)}\left(-\frac{\dot A}{A}-\frac{\dot B}{B}+\frac{\dot G}{G}\right).
\label{theta}
\end{eqnarray}

Next, the shear tensor
\begin{equation}
\sigma_{\alpha \beta}=\sigma_{(a)(b)}e^{(a)}_\alpha e^{(b)}_\beta=V_{(\alpha;\beta)}+a_{(\alpha}
V_{\beta)}-\frac{1}{3}\Theta h_{\alpha \beta}, \label{acc}
\end{equation}

 may be  defined through two independent tetrad components (scalars)  $\sigma_{(1)(1)}$ and $\sigma_{(2)(2)}$, 
which may be written in terms of the metric functions and their derivatives as (see \cite{1}):
\begin{eqnarray}
\sigma_{(1) (1)}&=&\frac{1}{3A}\left(\frac{\dot B}{B}-\frac{\dot C}{C}\right)\nonumber \\
&+&\frac{G^2}{3A\left(A^2 B^2 r^2 + G^2\right)}\left(\frac{\dot A}{A}+\frac{\dot B}{B}-\frac{\dot G}{G}\right),
 \label{sigmasI}
\end{eqnarray}
\begin{eqnarray}
\sigma_{(2)(2)}&=&\frac{1}{3A}\left(\frac{\dot B}{B}-\frac{\dot C}{C}\right)\nonumber \\ &+&\frac{2G^2}{3A\left (A^2 B^2 r^2 + G^2\right)}\left(-\frac{\dot A}{A}-\frac{\dot B}{B}+\frac{\dot G}{G}\right)
\label{sigmas}.
\end{eqnarray}

It is worth noticing that the shear tensor has no projection in the subspace $e^{(1)}_\alpha e^{(2)}_\beta$.

Finally,  for the vorticity tensor 
\begin{equation}
\Omega
_{\beta\mu}=\Omega_{(a)(b)}e^{(a)}_\beta
e^{(b)}_\mu,
\end{equation}
 we find that it is determined by a single basis component:
\begin{equation}
\Omega_{(1)(2)} = -\Omega_{(2)(1)}=-\Omega,
\label{omegan}
\end{equation}

where the scalar function $\Omega$ is given by
\begin{equation}
\Omega =\frac{G(\frac{G^\prime}{G}-\frac{2A^\prime}{A})}{2B\sqrt{A^2B^2r^2+G^2}}.
\label{no}
\end{equation}

Now, from the regularity conditions, necessary to ensure elementary flatness in the vicinity of  the axis of symmetry, and in particular at the center (see \cite{1n}, \cite{2n}, \cite{3n}), we should require  that as $r\approx 0$
\begin{equation}
\Omega=\sum_{n \geq1}\Omega^{(n)}(t,\theta) r^{n},
\label{sum1}
\end{equation}
implying, because of (\ref{no}) that in the neighborhood of the center
\begin{equation}
G=\sum_{n\geq 3}G^{(n)}(t, \theta) r^{n}.
\label{sum1}
\end{equation}

Beside the kinematical variables defined above, it would be convenient for our discussion  to introduce  the  ``specific velocities'', defined in \cite{qsa} (with the change of notation already mentioned):
\begin{equation}
V_{(1) (1)}=e^\alpha_{(1)} e^\beta_{(1)}(\sigma_{\alpha \beta}+\frac{1}{3}\Theta h_{\alpha \beta}+\Omega_{\alpha \beta}),
\label{vel1}
\end{equation}
\begin{equation}
V_{(2)(2)}=e^\alpha_{(2)} e^\beta_{(2)}(\sigma_{\alpha \beta}+\frac{1}{3}\Theta h_{\alpha \beta}+\Omega_{\alpha \beta}),
\label{vel2}
\end{equation}
\begin{equation}
V_{(3)(3)}=e^\alpha_{(3)} e^\beta_{(3)}(\sigma_{\alpha \beta}+\frac{1}{3}\Theta h_{\alpha \beta}+\Omega_{\alpha \beta}),
\label{vel3}
\end{equation}
\begin{equation}
V_{(1)(2)}=e^\alpha_{(1)} e^\beta_{(2)}(\sigma_{\alpha \beta}+\frac{1}{3}\Theta h_{\alpha \beta}+\Omega_{\alpha \beta}),
\label{vel4}
\end{equation}
which become, using (\ref{theta}), (\ref{sigmasI}), (\ref{sigmas}) and (\ref{omegan})
\begin{equation}
V_{(1)(1)}=\frac{1}{3}\left(3\sigma_{(1)(1)}+\Theta\right),\;
V_{(2)(2)}=\frac{1}{3}\left(3\sigma_{(2)(2)}+\Theta\right),
\label{vel7}
\end{equation}
\begin{equation}
V_{(3)(3)}=\frac{1}{3}\left(\Theta-3\sigma_{(1)(1)}-3\sigma_{(2)(2)}\right),\;
V_{(1)(2)}=-\Omega,
\label{vel8}
\end{equation}
satisfying
\begin{equation}
V_{(1)(1)}+V_{(2)(2)}+V_{(3)(3)}=\Theta.
\label{vel9}
\end{equation}
\\
The physical meaning of the above expressions becomes intelligible when we recall that the tensor $\sigma_{\alpha \beta}+\frac{1}{3}\Theta h_{\alpha \beta}+\Omega_{\alpha \beta}$ defines the proper time variation of the infinitesimal distance $\delta l$ between two neighboring points on the three-dimensional hypersurface (say $\Sigma$), orthogonal to the four velocity, divided by $\delta l$ (see \cite{qsa}) for details).
\\
\subsection{The electric and magnetic part of the Weyl tensor and the super--Poynting vector}
Let us now introduce the electric ($E_{\alpha\beta}$) and magnetic ($H_{\alpha\beta}$) parts of the Weyl tensor ( $C_{\alpha \beta
\gamma\delta}$),  defined as usual by
\begin{eqnarray}
E_{\alpha \beta}&=&C_{\alpha\nu\beta\delta}V^\nu V^\delta,\nonumber\\
H_{\alpha\beta}&=&\frac{1}{2}\eta_{\alpha \nu \epsilon
\rho}C^{\quad \epsilon\rho}_{\beta \delta}V^\nu
V^\delta\,.\label{EH}
\end{eqnarray}

The electric part of the Weyl tensor has only three independent non-vanishing components, whereas only two components define the magnetic part. Thus  we may  write these two  tensors, in terms of five  tetrad components  ($\mathcal{E}_{(1)(1)}, \mathcal{E}_{(2)(2)}, \mathcal{E}_{(1)(2)}, H_{(1)(3)}, H_{(3)(2)}$), respectively as:

\begin{widetext}
\begin{equation}
E_{\alpha\beta}=\left[\left(2\mathcal{E}_{(1)(1)}+\mathcal{E}_{(2)(2)}\right) \left(e^{(1)}_\alpha
e^{(1)}_\beta-\frac{1}{3}h_{\alpha \beta}\right)\right] +\left[\left(2\mathcal{E}_{(2)(2)}+\mathcal{E}_{(1)(1)}\right)\left (e^{(2)}_\alpha
e^{(2)}_\beta-\frac{1}{3}h_{\alpha \beta}\right)\right]+\mathcal{E}_{(2)(1)} \left(e^{(1)}_\alpha
e^{(2)}_\beta+e^{(1)}_\beta
e^{(2)}_\alpha\right), \label{E'}
\end{equation}
\end{widetext}
\noindent

and
\begin{equation}
H_{\alpha\beta}=H_{(1)(3)}\left(e^{(1)}_\beta
e^{(3)}_\alpha+e^{(1)}_\alpha
e^{(3)}_\beta \right)+H_{(2)(3)}\left(e^{(3)}_\alpha
e^{(2)}_\beta+e^{(2)}_\alpha
e^{(3)}_\beta \right)\label{H'}.
\end{equation}

Also, from  the Riemann tensor we may define  three tensors $Y_{\alpha\beta}$, $X_{\alpha\beta}$ and
$Z_{\alpha\beta}$ as

\begin{equation}
Y_{\alpha \beta}=R_{\alpha \nu \beta \delta}V^\nu V^\delta,
\label{Y}
\end{equation}
\begin{equation}
X_{\alpha \beta}=\frac{1}{2}\eta_{\alpha\nu}^{\quad \epsilon
\rho}R^\star_{\epsilon \rho \beta \delta}V^\nu V^\delta,\label{X}
\end{equation}
and
\begin{equation}
Z_{\alpha\beta}=\frac{1}{2}\epsilon_{\alpha \epsilon \rho}R^{\quad
\epsilon\rho}_{ \delta \beta} V^\delta,\label{Z}
\end{equation}
 where $R^\star _{\alpha \beta \nu
\delta}=\frac{1}{2}\eta_{\epsilon\rho\nu\delta}R_{\alpha
\beta}^{\quad \epsilon \rho}$  and $\epsilon _{\alpha \beta \rho}=\eta_{\nu
\alpha \beta \rho}V^\nu$.

The above tensors in turn, may be  decomposed, so that each of them is described through  four scalar functions known as structure scalars \cite{sc}. These are (see \cite{1} for details)
\begin{eqnarray}
Y_T&=&4\pi(\mu+3P), \qquad X_T=8\pi \mu, \label{ortc1}\\
Y_I&=&3\mathcal{E}_{(1)(1)}-12\pi \Pi_{(1)(1)},\quad X_I=-3\mathcal{E}_{(1)(1)}-12\pi \Pi_{(1)(1)}
\nonumber\\
Y_{II}&=&3\mathcal{E}_{(2)(2)}-12\pi \Pi_{(2)(2)},\quad X_{II}=-3\mathcal{E}_{(2)(2)}-12\pi \Pi_{(2)(2)}, \nonumber\\
Y_{III}&=&\mathcal{E}_{(2)(1)}-4\pi \Pi_{(2)(1)}, \quad X_{III}=-\mathcal{E}_{(2)(1)}-4\pi \Pi_{(2)(1)}.\nonumber
\end{eqnarray}
and

\begin{widetext}
\begin{equation}
Z_I=(H_{(1)(3)}-4\pi q_{(2)});\quad Z_{II}=(H_{(1)(3)}+4\pi  q_{(2)}); \quad Z_{III}=(H_{(2)(3)}-4\pi q_{(1)}); \quad  Z_{IV}=(H_{(2)(3)}+4\pi q_{(1)}). \label{Z2}
\end{equation}
\end{widetext}
From the above tensors, we may define  the super--Poynting
vector  by
\begin{equation}
P_\alpha = \epsilon_{\alpha \beta \gamma}\left(Y^\gamma_\delta
Z^{\beta \delta} - X^\gamma_\delta Z^{\delta\beta}\right),
\label{SPdef}
\end{equation}
where $\epsilon _{\alpha \beta \rho}=\eta_{\nu
\alpha \beta \rho}V^\nu$.

 In our case,  we may write:
\begin{equation}
 P_\alpha=P_{(1)}e^{(1)}_\alpha+P_{(2)}e^{(2)}_\alpha,\label{SP}
\end{equation}
with
\begin{widetext}
\begin{eqnarray}
P_{(1)} =
2H_{(2)(3)}\left(2{\cal E}_{(2)(2)}+{\cal E}_{(1)(1)}\right)+2H_{(1)(3)}{\cal E}_{(2)(1)}+ 32\pi^2 q_{(1)}\left[(\mu+P)+\Pi_{(1)(1)}\right] 
+32\pi^2 q_{(2)}\Pi_{(2)(1)} ,\nonumber
\\
P_{(2)}=-2H_{(1)(3)}\left(2{\cal E}_{(1)(1)}+{\cal E}_{(2)(2)}\right)-2H_{(2)(3)}{\cal E}_{(2)(1)}
+ 32\pi^2 q_{(2)}\left[(\mu+P)+\Pi_{(2)(2)}\right]+32\pi^2q_{(1)}\Pi_{(2)(1)} . \label{SPP}
\end{eqnarray}
\end{widetext}

In the theory of  the super--Poynting vector, a state of gravitational radiation is associated to a  non--vanishing component of the latter (see \cite{11p, 12p, 14p}). This is in agreement with the established link between the super--Poynting vector and the news functions \cite{5p}, in the context of the Bondi--Sachs approach \cite{7, 8}. 

We can identify two different contributions in (\ref{SPP}). On the one hand we have contributions from the  heat transport process. These are in principle independent of the magnetic part of the Weyl tensor, which explains why they  remain in the spherically symmetric limit.  Next we have contributions related to the gravitational radiation. These require, both, the electric and the magnetic part of the Weyl tensor to be different from zero.

\section{The heat transport equation}
In order to avoid the drawbacks generated by the standard (Landau--Eckart) irreversible thermodynamics \cite{17}, \cite{67}, (see \cite{63}-\cite{66} and references therein)  we shall need a transport equation derived from  a causal  dissipative theory   \cite{18, 19, 20, 21, 68, 69}. In this work we shall resort to 
M\"{u}ller-Israel-Stewart second
order phenomenological theory for dissipative fluids \cite{18, 19, 20, 21}). However, as we shall see, the main conclusions generated by our study are not dependent on the transport equation  chosen, as far as it is a causal one, i.e that it leads to a  Cattaneo type \cite{70} equation, leading
thereby to a hyperbolic equation for the propagation of thermal
perturbations.

Thus, the transport equation for the heat flux reads \cite{19, 20, 21, 64},
\begin{equation}
\tau h^\mu_\nu q^\nu _{;\beta}V^\beta +q^\mu=-\kappa
h^{\mu\nu}(T_{,\nu}+T a_\nu)-\frac{1}{2}\kappa T^2\left
(\frac{\tau V^\alpha}{\kappa T^2}\right )_{;\alpha}q^\mu,\label{qT}
\end{equation}

\noindent where $\tau$, $\kappa$, $T$ denote the relaxation time,
the thermal conductivity and the temperature, respectively.

Contracting (\ref{qT}) with $e^{(2)}_\mu$ we obtain
\begin{widetext}
\begin{eqnarray}
\frac{\tau}{A}\left(\dot q_{(2)}+A q_{(1)} \Omega\right)+q_{(2)}=-\frac{\kappa}{A}\left(\frac{G \dot T+A^2 T_{,\theta}}{\sqrt{A^2B^2r^2+G^2}}+A T a_{(2)}\right) -\frac{\kappa T^2q_{(2)}}{2}\left(\frac{\tau V^\alpha}{\kappa T^2}\right)_{;\alpha},\label{qT1n}
\end{eqnarray}
\end{widetext}
where  (\ref{no}), has been used

On the other hand,  contracting (\ref{qT}) with  $e^{(1)}_\mu$, we find

\begin{eqnarray}
\frac{\tau}{A}\left(\dot q_{(1)}-A q_{(2)} \Omega\right)+q_{(1)}=-\frac{\kappa}{B}\left(T^\prime+BTa_{(1)}\right)\nonumber \\
-\frac{\kappa T^2 q_{(1)}}{2}\left(\frac{\tau
V^\alpha}{\kappa T^2}\right)_{;\alpha}. \label{qT2n}
\end{eqnarray}

It is worth noticing  that the two equations above are coupled  through the vorticity. 

\section{Leaving the equilibrium}

We shall now take a snapshot of the system, just after it has abandoned the equilibrium. As mentioned before, by ``just after'' we mean on the smallest time scale, at which we can detect the first signs of dynamical evolution. 

The general ``philosophy'' of our approach consists of considering a fluid distribution which is in equilibrium (in the sense exposed in the Introduction), and assume that, for a reason which is not relevant for the discussion, at some initial time (say $t_0$) the system abandons such a state. Thus, at $t_0$ the  clock is put  to measure time, and we stop it as soon as we detect the first sign of dynamic evolution. The scale time under consideration is defined by the time interval measured by our clock. This is, so to speak, the ``philosophy'' of the approach. 

However, in practice we shall proceed slightly differently. Indeed, we are going to choose a given time scale,  which  we shall specify below. 
 Two possible results may then appear:
 \begin{itemize} 
\item No signs of dynamic evolution are observed within the choosen time scale 
\item  Such signs do appear, at such time scale. 
\end{itemize}
Of course in the case of the first result, we should have to enlarge our time scale.

Now, in the study of dissipative fluids, there are three fundamental time scales, each of which endowed with a distinct physical meaning, namely: the hydrostatic time (sometimes also called the hydrodynamic time), the thermal relaxation time and the thermal adjustment time (see \cite{astr1, astr2} for details).

The hydrostatic time is the typical time in which a fluid element reacts on a slight perturbation of hydrostatic equilibrium, it is basically of the order of magnitude of the time taken by a sound wave to propagate through the whole fluid distribution. 

The thermal relaxation time is the time taken by the system to return to the steady state in the heat flux (whether of thermodynamic equlibrium or not), after it has been removed from it.

 Finally, the thermal adjustment time is the time it takes a fluid element to adjust thermally to its surroundings.  It is, essentially, of the order of magnitude of the time  required for a significant change in the temperature gradients. From the above it is evident that the thermal adjustment time is, generally, larger than the thermal relaxation time.

We shall evaluate the system at a time scale  which is smaller than  the three  time scales described above. It should be emphasized that such a time scale is chosen heuristically. Thus, as mentioned before, if no sign of evolution could be detected within this time scale, it should be enlarged until these signs appear. However, as we shall see below, such signs do appear within the time scale under consideration.

The above comments  imply that:
\begin{itemize}
\item At the time scale at which we are observing the system, which is smaller than the hydrostatic time scale,  the kinematical quantities $\Omega (G), \Theta, \sigma_{(1) (1)},\sigma_{(2) (2)}$  as well as the ``velocities'' $V_{(1) (1)}, V_{(2) (2)},
V_{(3) (3)}, V_{(1) (2)}$ keep the same values they have in equilibirum, i.e. they are neglected (of course not so their time derivatives which are assumed to be small, say of order $O(\epsilon)$, where $\epsilon<<1$), but non--vanishing).

\item From (\ref{esc6K}) (\ref{esc61L}) (Eqs. B6, B7 in \cite{1}), it follows at once that the heat flux vector should also be   neglected (once again, not so its time derivative). The vanishing of the flux vector also follows at once from the fact that the time scale under consideration is smaller than the relaxation time.
\item From the above conditions it follows at once that  first order time derivatives of the metric variables $A, B, C$ can be neglected.

\end{itemize}

Then, we have for the four acceleration
\begin{equation}
a_{(1)}= \frac {A^\prime }{AB };\quad a_{(2)}=\frac{1}{Br}\left(\frac {A_{,\theta}}{A}+\frac{\dot G}{A^2}\right).
\label{accitem}
\end{equation}

Also, from the conditions above and (\ref{theta}, \ref{sigmasI}, \ref{sigmas}, \ref{no}, \ref{vel7}, \ref{vel8}), it follows that
\begin{equation}
\dot \Theta=\frac{1}{A}\left(\frac{2\ddot B}{B}+\frac{\ddot C}{C}\right),\quad\dot \sigma_{(1) (1)}=\dot \sigma_{(2) (2)}\equiv  \dot {\bar \sigma}=\frac{1}{3A}\left(\frac{\ddot B}{B}-\frac{\ddot C}{C}\right),
\label{ds}
\end{equation}
\begin{equation}
\dot \Omega=\frac{1}{AB^2r}\left(\frac{\dot G^{\prime}}{2}-\frac{\dot G A^{\prime}}{A}\right),
\label{vornn}
\end{equation}
and
\begin{equation}
V_{(1) (1)}=V_{(2) (2)}\equiv V, \quad \dot V=\frac{\ddot B}{AB}, \quad \dot V_{(3) (3)}=\frac{\ddot C}{AC}.
\label{dsv}
\end{equation}
Now, at thermal equilibrium,  when the heat flux vanishes, the Tolman conditions for thermal equilibrium  \cite{Tolman}
\begin{equation}
(TA)^\prime=(TA)_{,\theta}=0,
\label{tol}
\end{equation}
are valid. 

Therefore  just after the system leaves the equilibrium, at a time scale which is smaller than the thermal adjustment  time and the thermal relaxation time, the equations (\ref{tol}) are still valid, even though the system starts to leave the thermal equilibrium.  This is so because of the fact that our time scale is smaller than the relaxation time, and therefore the temperature gradients have the same values they had in equilibrium. However, the fulfillment of (\ref{tol}) is not enough to ensure the vanishing of $\dot q_{(2)}$, due to the appearance of a $\dot G$ term in (\ref{qT1n}) (through $a_{(2)}$), which   eventually would lead to the breaking of  the thermal equilibrium in the meridional direction (at later time).

Thus, the evaluation of (\ref{qT2n})  and (\ref{qT1n}) just after leaving the equilibrium, produces respectively

\begin{eqnarray}
\dot q_{(1)}=0 ,\label{qT1nn}
\end{eqnarray}
and

\begin{eqnarray}
\tau \dot q_{(2)}=-\frac{\kappa AT_{,\theta}}{Br}-\kappa ATa_{(2)},
 \label{qT2nnn}
\end{eqnarray}
or, using (\ref{tol})
\begin{eqnarray}
\tau \dot q_{(2)}=-\frac{\kappa T \dot G}{ABr}
. \label{qT2nn}
\end{eqnarray}

Therefore, at the very beginning of  the evolution, the dissipative process starts with contributions along the $e^{(2)}_\mu$ (meridional) direction.

We shall now turn to fluid variables ($\mu, P, \Pi_{(1)(1)}, \Pi_{(2)(2)}, \Pi_{(2)(1)}$). 
Using MAPLE we shall calculate the components of the Einstein tensor $G_{\alpha \beta}$ and evaluate them just after the system leaves the equilibrium. At this time scale, this tensor have three types of terms: On the one hand, terms with first time derivatives of the metric functions $A, B, C$, which are  are set to zero, next, there are terms  that neither contain $G$, nor first time derivatives of  $A, B, C$,  these  correspond to the expression in equilibrium, finally, there are terms with first time derivatives of $G$ and/or  second time derivatives of $A, B, C$,  which of course are not neglected. Then using (\ref{eisp}, \ref{7P}, \ref{2n}, \ref{2nbis}) and the  Einstein equations , 
\begin{equation}
G_{\alpha \beta}=-8\pi T_{\alpha \beta},
\label{einst}
\end{equation}
 we obtain
\begin{equation}
8\pi \mu=8\pi\mu_{(eq)},
\label{moeq}
\end{equation}
\begin{equation}
8\pi P=8\pi P_{(eq)}-\frac{2}{3A}\dot \Theta+\frac{2}{3A^2B^2r^2}\left(\dot G_{,\theta}+\dot G\frac{C_{,\theta}}{C}\right),
\label{Poeq}
\end{equation}
\begin{equation}
8\pi \Pi_{(1) (1)}=8\pi \Pi_{(1)(1) (eq)}+\frac{\dot {\bar \sigma}}{A}+\frac{1}{3A^2B^2r^2}\left[\dot G_{,\theta}-\dot G\left(\frac{3B_{,\theta}}{B}-\frac{C_{,\theta}}{C}\right)\right],
\label{PiIoeq}
\end{equation}
\begin{eqnarray}
8\pi \Pi_{(2)(2)}=8\pi \Pi_{(2)(2) (eq)}+\frac{\dot {\bar \sigma}}{A}\nonumber \\+\frac{1}{3A^2B^2r^2}\left[-2\dot G_{,\theta}+\dot G\left(\frac{3B_{,\theta}}{B}+\frac{C_{,\theta}}{C}\right)\right],
\label{PiIIoeq}
\end{eqnarray}
\begin{equation}
8\pi \Pi_{(2)(1)}=8\pi \Pi_{(1)(1)(eq)}-\frac{\dot \Omega}{A}+\frac{\dot G}{A^2B^2r}\left[\frac{(Br)^\prime}{Br}-\frac{A^\prime}{A}\right],
\label{PiKLoeq}
\end{equation}
where $eq$ stands for the value of the quantity at equilibrium.

Now, from (\ref{moeq}) it follows at once that the energy density, after leaving the equilibrium, at the time scale considered here, has the same value it had in equilibrium. Then since there should be a generic equation of state relating the energy density with the isotropic pressure, it is reasonable to assume that at the  time scale under consideration we have $P=P_{(eq)}$, and following this line of arguments it would be also reasonable to assume $ \Pi_{(1)(1)}=\Pi_{(1)(1) (eq)},  \Pi_{(2)(2)}=\Pi_{(2)(2)(eq)}$. 

Once again, it is  important to remark that such  assumptions are purely heuristic. Therefore if it would happen that as a consequence of their imposition, we detect no signs of evolution (at the time scale under consideration), we should relax them  and enlarge our time scale, until these signs become observable. However this is not the case. Indeed, from these latter conditions and (\ref{no}, \ref{Poeq}, \ref{PiIoeq}, \ref{PiIIoeq}), it follows at once that:
\begin{equation}
\dot G=B^2f(t,r),
\label{bnf}
\end{equation}
\begin{equation}
\dot {\bar \sigma}=-\frac{f(t,r)}{3Ar^2}\left(\frac{C_{,\theta}}{C}-\frac{B_{,\theta}}{B}\right),
\label{bnf1s}
\end{equation}
\begin{equation}
\dot\Theta=\frac{f(t,r)}{Ar^2}\left(\frac{2B_{,\theta}}{B}+\frac{C_{,\theta}}{C}\right),
\label{bnf1}
\end{equation}
\begin{equation}
\dot \Omega=\frac{f(t,r)}{Ar}\left(\ln{\frac{B\sqrt{f}}{A}}\right)^\prime,
\label{bnf1omega}
\end{equation}
where $f(t,r)$ is an arbitrary function of its arguments.

Two comments are in order at this point:
\begin{itemize}
\item Because of (\ref{sum1}) it is obvious that $f=\sum_{n\geq 3}f^{(n)}(t) r^{n}$ in the neighborhood of the center.
\item Observe that $f$ controls the evolution of $G (\Omega), \Theta$ and $\bar \sigma$.
\end{itemize}

The situation is quite different for  the scalar $\Pi_{(2)(1)}$. In fact, as we shall see, we cannot  assume that $\Pi_{(2)(1)}=\Pi_{(2)(1)(eq)}$.

Indeed, because of (\ref{PiKLoeq}), to assume that $\Pi_{(2)(1)}=\Pi_{(2)(1)(eq)}$, amounts  to impose the condition
\begin{equation}
\frac{\dot \Omega}{A}=\frac{\dot G}{A^2B^2r}\left[\frac{(Br)^\prime}{Br}-\frac{A^\prime}{A}\right],
\label{PiKLoeqb}
\end{equation}
which together with (\ref{no}) produces
\begin{equation}
\dot G=B^2r^2g(t,\theta),
\label{bnfbis}
\end{equation}
where $g$ is an arbitrary function of its arguments. But, (\ref{bnfbis}) clearly violates the regularity condition  (\ref{sum1}), close to the center. Accordingly, at the time scale under consideration we have $\Pi_{(2)(1)}\neq\Pi_{(2)(1)(eq)}$, more precisely
\begin{equation}
8\pi \Pi_{(2)(1)}=8\pi \Pi_{(2)(1) (eq)}+\frac{f(t,r)}{2A^2r}\left(\ln{\frac{r^2}{f}}\right)^\prime.
\label{PiKLoeqbb}
\end{equation}
Thus we see that, after leaving the equlibrium, at the time scale under consideration,  the  energy density, the isotropic pressure and the $(1)(1)$ and the  $(2)(2)$ tetrad components of the anisotropic tensor may be assumed to keep the values they have in equilibrium. However for the transverse tension $\Pi_{(2)(1)}$ the situation is different, and the first signs of the dynamic regime are already present in this tetrad  component of the anisotropic tensor, at our time scale.

Using MAPLE we can also easily  calculate the scalars defining the electric part of the Weyl tensor, after the system leaves the equlibrium, we obtain:
\begin{widetext}
\begin{equation}
 {\cal E}_{(1)(1)}={\cal E}_{(1)(1)(eq)}-\frac{\dot {\bar \sigma}}{2A}-\frac{1}{6A^2B^2r^2}\left[\dot G_{,\theta}-\dot G\left(\frac{3B_{,\theta}}{B}-\frac{C_{,\theta}}{C}\right)\right],
\label{EIoeq}
\end{equation}

\begin{equation}
{\cal E}_{(2)(2)}={\cal E}_{(2)(2) (eq)}-\frac{\dot {\bar \sigma}}{2A}+\frac{1}{6A^2B^2r^2}\left[2\dot G_{,\theta}-\dot G\left(\frac{3B_{,\theta}}{B}+\frac{C_{,\theta}}{C}\right)\right]
\label{EIIoeq}
\end{equation}
\begin{equation}
{\cal E}_{(2)(1)}={\cal E}_{(2)(1) (eq)}+\frac{\dot \Omega}{2A}-\frac{\dot G}{2A^2B^2r}\left[\frac{(Br)^\prime}{Br}-\frac{A^\prime}{A}\right].
\label{EKLoeq}
\end{equation}
\end{widetext}
Using (\ref{PiKLoeq}), (\ref{bnf}) and (\ref{bnf1}) in (\ref{EIoeq}), (\ref{EIIoeq})  and (\ref{EKLoeq}), it follows at once that
\begin{equation}
{\cal E}_{(1)(1)}={\cal E}_{(1)(1) (eq)},\quad {\cal E}_{(2)(2)}={\cal E}_{(2)(2) (eq)},\quad {\cal E}^{oeq}_{(2)(1)}=-4\pi\Pi^{oeq}_{(2)(1)},
\label{e,pies}
\end{equation}
which impliy, because of (\ref{ortc1})
\begin{equation}
X_I=X_{I (eq)},\quad X_{II}=X_{II (eq)},\quad X_{III}=X_{III (eq)},
\label{xs}
\end{equation}
and
\begin{equation}
Y_I=Y_{I (eq)},\quad Y_{II}=Y_{II (eq)},\quad Y^{oeq.}_{III}=-8\pi\Pi^{oeq.}_{(2)(1)},
\label{ys}
\end{equation}
where $oeq$ stands for the value of the quantity  ``out of equilibrium'', and as it follows at once from (\ref{PiKLoeqbb})
\begin{equation}
8\pi \Pi^{oeq}_{(2)(1)}=\frac{f(t,r)}{2A^2r}\left(\ln{\frac{r^2}{f}}\right)^\prime.
\label{PiKLoeqbbc}
\end{equation}

Let us now analyze the ``generalized Euler equations``  (\ref{ec2}) (Eq. A7 in \cite{1}), derived from the ``conservation laws`` ($T^{\mu \nu}_{;\nu}=0$). Evaluated within the time scale under consideration, these are the equations (\ref{ge3bn}) and (\ref{ge2b}) in the Appendix:

Observe that these two equations  have the  ``Newtonian'' form 
\begin{equation}
 Mass \; density \times Acceleration=Force,
\label{Newton}
\end{equation}
\\
and where we can  clearly  identify the ``effective inertial mass density'' as the factors multiplying $\dot V$ and $\dot V_{(3)(3)}$. Also, it is worth noticing that the first term in the right hand side of (\ref{ge3bn}), and the first term   in the right hand side of (\ref{ge2b}),  represent the ``gravitational force''.  This is  in agreement with  the equivalence principle, according to which, the ``effective inertial mass density'' equals the ``passive gravitational mass density'' (the factor multiplying the square brackets in (\ref{ge3bn}) and  (\ref{ge2b})).

We observe that, according to (\ref{ge3bn}) and (\ref{ge2b}) there are two different  ``effective inertial mass densities'', depending on the anisotropy of the fluid. This is a clear reminiscence of  the situation appearing in relativistic dynamics, where a moving particle offers different inertial resistances to the same force, according to whether it is subjected to that force longitudinally or transversely.

Finally, replacing $\dot q_{(2)}$, by its expression from (\ref{qT2nnn}), into   (\ref{ge2b}) we obtain
\begin{widetext}
\begin{eqnarray}
&&\left(\mu+P+\Pi_{(2)(2)}\right)\left[1-\frac{\kappa T}{\tau(\mu+P+\Pi_{(2)(2)})}\right]\dot V_{(3)(3)}=-\left(\mu+P+\Pi_{(2)(2)}\right)\left[1-\frac{\kappa T}{\tau(\mu+P+\Pi_{(2)(2)})}\right]\left[4\pi A\left(P+2\Pi_{(1)(1)}+2\Pi_{(2)(2)}\right)\right.
\nonumber \\
&&\left.-\frac{AC^\prime a_{(1)}}{BC}\right] + {\rm force \, and \, dissipative \, terms}.
\label{ge2bb}
\end{eqnarray}
\end{widetext}

This last equation illustrates the well known decreasing of the inertial mass density (and consequently, of the passive gravitational mass density) associated to thermal effects, which was discovered   in \cite{22}, and that has been shown to appear in a great variety of scenarios (see \cite{23, 40, 41, 24, 25, 26, 27, 28, 29, 30, 42, 43, 44} and references therein).

Next, observe that by evaluating  the physical variables  out of equlibrium, we may obtain
\begin{equation}
\dot V=\frac{B_{,\theta}\dot G}{AB^3r^2}=\frac{B_{,\theta} f}{Ar^2B},
\label{vcon1}
\end{equation}
\begin{equation}
\dot V_{(3)(3)}=\frac{C_{,\theta}\dot G}{ACB^2r^2}=\frac{C_{,\theta }f}{Ar^2C},
\label{vcon2}
\end{equation}
from where it is apparent  that $f$ controls the evolution of the  different ``velocities''.

We can now turn to the equations (B1, B3) and (B5) in \cite{1}.They  describe the evolution of $\Theta$, $\bar \sigma$ and $\Omega$, and  using (\ref{bnf1s}, \ref{bnf1}, \ref{bnf1omega}) they become identities. On the other hand (B4) becomes an identity when using (\ref{bnf}, \ref{PiKLoeqbb}, \ref{e,pies}).

Next we have the equations (\ref{esc6K}), (\ref{esc61L}) (B6, B7 in \cite{1}), which from the all  results obtained above become identities, whereas the equations (B8) and (B9) imply
\begin{equation}
H_{(1)(3)}=H_{(3)(2)}=0,
\label{mag}
\end{equation}
of course their time derivatives do not vanish, as we shall see below.

Equations (B10--B13) in \cite{1} describe the evolution of the structure scalars $X_I, X_{II}, X_{III}$. It is a simple matter to check that within the time scale considered here $\dot X_I=\dot X_{II}=\dot X_{III}=0$. Also, it is a simple matter to see that equations (B14--B16) in \cite{1} do not provide any additional information.

Finally, the equations (\ref{esc10SK}) and (\ref{esc10SL}) (B17, B18 in \cite{1}) describe the evolution of the magnetic part of the Weyl tensor in terms of the function $f(t,r)$, more specifically, these equations become:
\begin{eqnarray}
\dot H_{(1)(3)}&=&\frac{f}{4ABr}\left[\frac{f^\prime}{rf}-\frac{f^{\prime \prime}}{f}-\left(\frac{2}{r}-\frac{f^\prime}{f}\right)\left(\frac{A^\prime}{A}-\frac{2B^\prime}{B}+\frac{C^\prime}{C}\right)\right]\nonumber \\&+&\frac{fB\left(2 {\cal E}_{(1)(1)}+ {\cal E}_{(2)(2)}\right)}{Ar},
\label{eh1}
\end{eqnarray}
\begin{eqnarray}
\dot H_{(3)(2)}&=&\frac{f}{4ABr^2}\left(\frac{2}{r}-\frac{f^\prime}{f}\right)\left(\frac{A_{,\theta}}{A}-\frac{2B_{,\theta}}{B}+\frac{C_{,\theta}}{C}\right)\nonumber \\&+&\frac{fB{\cal E}_{{(2)(1)}(eq)}}{Ar},
\label{eh2}
\end{eqnarray}
from which  it is evident that the evolution of the magnetic part of the Weyl tensor is fully controlled by the function $f$.

\section{Conclusions}
We have carried out an exhaustive analysis of axially symmetric fluid distributions, just after its departure from equilibrium, at the smallest time scale at which we can detect signs of dynamical evolution. 

As our main result, we have found that the evolution of all variables is controlled by a single function $f$, which we call the fluid news function, in analogy with the Bondi's news function. Indeed,  if anything happens at all at the source leading to changes in the field, it can only do so through the function $f$, and viceversa, exactly as it appears from the analysis of the spacetime outside the source (Bondi). However, an important difference between these two functions must be emphasized, namely: our function $f$ controls the evolution only within the time scale considered here, a limitation which does not apply to the Bondi's news function  (see below for a deeper discussion on this point).

Among all the physical variables, there are two, which  play a significant role in the departure from equilibrium. On the one hand,  it is the heat flow  along the $e^\mu_{(2)}$ direction, the one which shall appear first. On the other hand, it is also remarkable that it is the  tetrad  component of the anisotropic tensor, in the subspace spanned by the tensor $e^\mu_{(2)}e^\nu_{(1)}$, the one which shows the first indications of the departure from equilibrium.

It is worth mentioning, that at the time scale used here, there is not gravitational radiation, as it follows at once from (\ref{SPP}). Thus, the emission of gravitational waves is an event which occurs at later times. This fact becomes intelligible at the light of the following comments.

For a second order phenomenological theory for dissipative fluids we obtain from Gibbs equation and conservation equations (see \cite{64, 28} for details):
\begin{widetext}
\begin{eqnarray}
T S^{\alpha}_{;\alpha} =
- q^{\alpha} \left[ h^\mu_{\alpha} (\ln{T })_{,\mu} +
V_{\alpha;\mu} V^\mu
 + \beta_{1} q_{\alpha;\mu} V^\mu+
\frac{T}{2} \left(
\frac{\beta_{1}}{T}V^{\mu}\right)_{;\mu}q_{\alpha}\right],
\label{diventropia}
\end{eqnarray}
\end{widetext}
where $S^\alpha$ is the entropy four--current,  and $\beta_1=\frac{\tau}{\kappa T}$.

From which it becomes evident that at the time scale under consideration  $S^\alpha_{;\alpha}=0$.

We recall that in the above expression, terms involving couplings of heat flux  to the vorticity, vanish at the time scale under consideration. Also,  we have excluded  shear  and bulk viscosity contributions in (\ref{diventropia}). The fact is that these absent terms are proportional to the shear tensor, the expansion scalar, terms quadratic in  the bulk viscosity pressure, terms proportional to   the bulk viscosity pressure multiplied by its time derivative,  and  terms proportional to the anisotropic stress tensor associated to the shear viscosity multiplied  by itself, or by its time derivative (see Eq.(2.20) in \cite{64}), (we recall that the anisotropic stress tensor may, but does not need to, be related to viscosity effects, since it may be sourced by many other physical phenomena.  Thus, for example it may be different from zero for a static configuration). Of course, within the time scale used here, all these terms vanish. However, it should be clear that in the study of any specific astrophysical scenario,  these dissipative phenomena  may be present and might play an important role in the detailed description of the structure and evolution of the object (at a time scale larger than the one considered here).  

Thus, within our time scale, our  observers do not  detect a real (entropy producing) dissipative process. But  as it was already pointed out in  the seminal Bondi's paper  on gravitational radiation(see section 6 in \cite{7}),  in the absence of dissipation, the system is not expected  to radiate (gravitationally) due to the reversibility of the equation of state, at variance with  the fact that  radiation is an irreversible process (see also \cite{hetaln} for a further discussion on this point). 

Therefore,  it is obvious that, in the presence  of gravitational radiation,  an entropy generator factor should also be present in the description of the source.  But  as we have just seen, such a factor does not  appear within the time scale under consideration. Accordingly it is reasonable, not to detect gravitational radiation at that same time scale. 

The reversibility of  the evolution, at the time scale under consideration,  implied by the above comments, could also be inferred from a simple inspection of (\ref{qT2nn}), (\ref{bnf}), (\ref{bnf1s}), (\ref{bnf1}), (\ref{bnf1omega}), (\ref{vcon1}), (\ref{vcon2}), (\ref{eh1}), (\ref{eh2}).

 Indeed, it  results at once from these equations, that if the function $f$ is different from zero until some time, and vanishes afterwards  (always within the  time scale under consideration), the system will  turn back  to equilibrium, without ``remembering'' to have been out of it previously.

 In other words, the fluid news function, unlike the Bondi's news function, is the precursor of, (appears before),  the dissipative process related to the emission of gravitational radiation, and should be different from zero until such emission starts. 

In relation with the point above, another comment is in order: in \cite{5p} the link between radiation and vorticity was put in evidence (see also \cite{vorh}), more specifically it was explicitly assumed that such a link was a causal one (the title of \cite{5p} is: ``Why does gravitational radiation produce vorticity?''), i.e. it was assumed that radiation precedes the appearance of vorticity. However as we have just shown,  both the magnetic part of the Weyl tensor, and $\Omega$ vanish at the time scale under consideration, whereas their first time derivatives do not vanish at that same time scale, suggesting that both phenomena (radiation and vorticity) occur essentially simultaneously.

An interesting particular case is represented by the situation appearing if we impose that the system was initially spherically symmetric (besides  of being in equilibrium), and assume that it remains spherically symmetric afterwards. In such  a case, it is obvious that we must have $\dot G=0$, implying  that departures from equilibrium (dynamic and thermal) only occur if Tolman's conditions (\ref{tol}), are violated. However, since the system was initially at equilibrium, such a violation may only happen at time scales larger that the thermal adjustment time. In other words, departures from equilibrium, keeping the spherical symmetry, take place at time scales larger than the corresponding to the, general, non--spherical case. Observe that in the purely spherically symmetric case the assumptions $P=P_{(eq)}$,  $ \Pi_{(1)(1)}=\Pi_{(1)(1)(eq)},  \Pi_{(2)(2)}=\Pi_{(2)(2)(eq)}$ do not hold (since we have to enlarge the time scale in order to observe the first signs of evolution), and of course the onset of evolution is not controlled by the function $f$ as defined by (\ref{bnf}).

We would like to emphasize  the appearance of the thermal effect leading to  a decreasing of the effective inertial mass density. In this respect, it is worth stressing that the first term on the left, and the $Ta_\nu$ term on the right, of  (\ref{qT}), are directly responsible for the decreasing in the effective inertial mass density. The former  should be present in any causal theory of dissipation, whereas the latter is just an expression of the ``inertia'' of heat already pointed out by Tolman \cite{Tolman}. 

Therefore any hyperbolic, relativistic  dissipative theory yielding a
Cattaneo-type equation in the non-relativistic limit, is expected to give a result similar to the one obtained here. The possible consequences  of this effect on the outcome of gravitational collapse  have been discussed in some detail in \cite{25, 26}. It is also worth noticing that such an effect appears already at the earliest stages of the non--equilibrium (though only along the $V_{(3)(3)}$ direction).

Finally we would like to conclude with the following remark:   In the stationary case  one may have a steady rotation around the symmetry axis, leading to non vanishing (time independent) vorticity $\Omega_{\mu\nu}\ne 0$, which of course may be compatible with thermal equilibrium. In this case the spacetime outside the source is described by a metric of the Lewis-Papapetrou family (e.g. Kerr) which as we know admits vorticity in the congruence of the world  line of observers (the line element is non-diagonal). The vorticity of the source produces the vorticity in the exterior spacetime.  However,  in the static situation (the one considered here) you have no vorticity at the outside, which is described by a metric of the Weyl family (e.g.Curzon, Erez-Rosen, etc).  In this latter case (non stationary) we must have $\Omega_{\mu\nu}= G=0$ since the metric is diagonal. Since you have no vorticity outside (no frame dragging), you should not expect to have vorticity in the source (see \cite{static} for a discussion on this case).

 This last result may be obtained in a  more rigourous way, by evaluating  (\ref{esc6K}) and (\ref{esc61L}) in the static case and thermal equilibrium (assuming $\Omega\neq 0$). Then after some lengthy but simple calculations, and using the regularity condition (\ref{sum1}), one obtains $\Omega=0$. Thus there is no vorticity associated to the static case. This brings out the difference between the steady vorticity of the stationary case and the vorticity considered here.

Also, the result above, shows that  vorticity and heat flux are inherently coupled. This fact was already emphasized in \cite{1}.

\begin{acknowledgments}
L.H. thanks  Departament de F\'isica at the  Universitat de les  Illes Balears, for financial support and hospitality. ADP  acknowledges hospitality of the
 Departament de F\'isica at the  Universitat de les  Illes Balears. L.H and J.O. acknowledge financial support from the Spanish Ministry of Science and Innovation (grant FIS2009-07238) and  Fondo Europeo de Desarrollo Regional (FEDER) (grant FIS2015-65140-P) (MINECO/FEDER)
.
\end{acknowledgments}

\appendix 
\section{Some basic equations}
In what follows we shall deploy only those  equations of the formalism which are required for our discussion. The  whole set of the equations can be found in \cite{1}.

The conservation law $T^\alpha _{\beta;\alpha}=0$ leads to the following equations (Eqs. A6, A7 in \cite{1}):
\begin{widetext}
\begin{equation}
\mu _{;\alpha}V^\alpha +(\mu+P)\Theta +\left(2\sigma _{(1)(1)}+\sigma _{(2)(2)}\right)\Pi_{(1)(1)}+\left(2\sigma_{(2)(2)}+\sigma _{(1)(1)}\right)\Pi_{(2)(2)}+ q^{\alpha}_{;\alpha} + q^\alpha a_\alpha =0,\label{esc1}
\end{equation}

\begin{equation}
(\mu+P)a_\alpha+h_{\alpha}^\beta\left
(P_{;\beta}+\Pi_{\beta;\mu}^\mu+q_{\beta;\mu}V^\mu\right )+\left(
\frac{4}{3}\Theta h_{\alpha \beta}+\sigma_{\alpha
\beta}+\Omega_{\alpha\beta}\right )q^\beta=0.\label{ec2}
\end{equation}
\end{widetext}
The first of these equations is the ``continuity'' equation, whereas the second one is the ``generalized Euler'' equation.  

This last equation has two components, which, within the time scale under consideration may be written as:
\begin{widetext}
\begin{eqnarray}
\left(\mu+P+\Pi_{(1)(1)}\right)\dot V=-\left(\mu+P+\Pi_{(1)(1)}\right)\left[4\pi A\left(P-2\Pi_{(1)(1)}\right) -\frac{AB_{,\theta} a_{(2)}}{B^2r}\right] + ``force \quad terms'',
\label{ge3bn}
\end{eqnarray}
\end{widetext}
and
\begin{widetext}
\begin{eqnarray}
\left(\mu+P+\Pi_{(2)(2)}\right)\dot V_{(3)(3)}&=&-\left(\mu+P+\Pi_{(2)(2)}\right)\left[4\pi A\left(P+2\Pi_{(1)(1)}+2\Pi_{(2)(2)}\right)-\frac{AC^\prime a_{(1)}}{BC}\right]
-\frac{AC_{,\theta}}{BCr}\left[\frac{\dot q_{(2)}}{A}\right],\nonumber \\ &+ &``force \quad terms'',
\label{ge2b}
\end{eqnarray}
\end{widetext}
where by ``force terms'' we  denote different terms containing pressure gradients and anisotropic stresses.

Next, from the Ricci identities we have (Eqs. B6, B7 in \cite{1})
\begin{widetext}
\begin{eqnarray}
\frac{2}{3B}\Theta _{,r}-\Omega _{;\mu}e^\mu_{(2)}+\Omega \left(e^{(2)}_{\beta ;\mu}e^\mu_{(1)} e^\beta_{(1)}-e^\mu_{(2);\mu}\right)+\sigma _{(1)(1)} a_{(1)}-\Omega a_{(2)}-\sigma_{(1)(1);\mu}e^\mu_{(1)}\nonumber
\\
-\left(2\sigma _{(1)(1)}+\sigma _{(2)(2)}\right)\left(e^\mu_{(1) ;\mu}-\frac{a_{(1)}}{3}\right)-\left(2\sigma _{(2)(2)}+\sigma _{(1)(1)}\right)\left(e^{(2)}_{\beta ;\mu}e^\mu_{(2)} e^\beta_{(1)}-\frac{a_{(1)}}{3}\right)=8\pi
q_{(1)},\label{esc6K}
\end{eqnarray}

\begin{eqnarray}
\frac{1}{3\sqrt{A^2B^2r^2+G^2}}\left(\frac{2G}{A}
\Theta_{,t}+2A\Theta _{,\theta}\right )+a_{(2)} \sigma _{(2)(2)}+\Omega _{;\mu}e^{\mu}_{(1)}+\Omega \left(e^{\mu}_{(1) ;\mu}+e^{\mu}_{(2)} e^{\beta}_{(1)} e^{(2)}_{\beta;\mu}\right)+\Omega a_{(1)}-\sigma_{(2)(2);\mu}e^{\mu}_{(2)} \nonumber\\
+\left(2\sigma _{(1)(1)}+\sigma_{(2)(2)}\right)\left(e^{(2)}_{\beta;\mu} e^\beta_{(1)} e^\mu_{(1)}+\frac{a_{(2)}}{3}\right)-\left(2\sigma _{(2)(2)}+\sigma_{(1)(1)}\right)\left(e^\mu_{(2) ;\mu}-\frac{a_{(2)}}{3}\right)=8\pi q_{(2)}.
\label{esc61L}
\end{eqnarray}

Finally, from the Bianchi identities, the following two equations describing the evolution of the magnetic part of the Weyl tensor, are obtained (Eqs. B17, B18 in \cite{1}).

\begin{eqnarray}
-2 a_{(2)} \mathcal{E}_{(1)(1)}+2a_{(1)}\mathcal{E}_{(2)(1)}-E^\delta _{2;\delta}
e^2_{(2)}-\frac{AY_{I,\theta}}{3\sqrt{A^2B^2r^2+G^2}}+\frac{Y_{III,r}}{B} \nonumber
\\
-\left [\frac{1}{3}(2Y_I+Y_{II})e^{(1)}_{\beta;\delta}+\frac{1}{3}(2Y_{II}+Y_I)e^\nu_{(1)} e^{(2)}_{\nu;\delta}e^{(2)}_\beta+Y_{III}(e^{(2)}_{\nu;\delta} e^\nu_{(1)} e^{(1)}_\beta+e^{(2)}_{\beta;\delta})\right ]\epsilon ^{\gamma\delta \beta}e^{(3)}_\gamma\nonumber
\\
+H_{(3)(1),\delta}V^\delta +H_{(3)(1)}\left(\Theta+\sigma _{(2)(2)}-\sigma _{(1)(1)}\right)+\Omega H_{(2)(3)} =-\frac{4\pi}{3}\mu
_{,\theta}e^2_{(2)}+12\pi \Omega q_{(1)}+\frac{4\pi q_{(2)}}{3}\left(3\sigma _{(1)(1)}+\Theta\right),\label{esc10SK}
\end{eqnarray}

\begin{eqnarray}
2a_{(1)}\mathcal{E} _{(2)(2)}-2a_{(2)}\mathcal{E}_{(2)(1)}+E^{\delta}_{\beta;\delta}e^\beta_{(1)}
+\frac{Y_{II,r}}{3B}-\frac{A Y_{III,\theta}}{\sqrt{A^2B^2r^2+G^2}}\nonumber
\\
-\left [-\frac{1}{3}(2Y_I+Y_{II})e^{(2)}_{\nu;\delta}e^\nu_{(1)} e^{(1)}_\beta+\frac{1}{3}(2Y_{II}+Y_I)e^{(2)}_{\beta;\delta}+Y_{III}(e^{(1)}_{\beta;\delta}-e^\nu_{(1)} e^{(2)}_\beta
e^{(2)}_{\nu;\delta})\right ]\epsilon ^{\gamma\delta \beta}e^{(3)}_\gamma\nonumber
\\
+H_{(2)(3),\delta}V^\delta +H_{(2)(3)}\left(\Theta+\sigma _{(1)(1)}-\sigma _{(2)(2)}\right)-\Omega
H_{(1)(3)}=\frac{4\pi}{3}\mu_{,\beta}e^\beta_{(1)}-\frac{4\pi q_{(1)}}{3}\left(3\sigma _{(2)(2)}+\Theta\right)+12\pi \Omega q_{(2)}.\label{esc10SL}
\end{eqnarray}
\end{widetext}

\end{document}